# Aesthetic Motivation in Quantum Physics: Past and Present

*Henrik Zinkernagel*

This essay explores the relations between aesthetics and motivation, primarily in quantum physics, focusing on the notions of play, beauty, and the joy of insight. The motivating role of these notions is examined both historically among the quantum pioneers and contemporarily, based on a series of interviews, among physicists associated with the Niels Bohr Institute in Copenhagen.

## 1. Physics and Aesthetics

It is common for physicists to employ aesthetic language. They may use terms such as beauty, harmony and wonder—not only to describe their theories, explanations, and experimental results, but also to express their motivations or passionate involvement in the scientific process. Aesthetic language and considerations were also present among the pioneers of quantum physics. For instance, Werner Heisenberg once noted in retrospect that his early work in quantum mechanics contrasted with Niels Bohr's due to a certain "aesthetic judgment" regarding an abstract mathematical scheme that exerted a "magical attraction" on Heisenberg but not on Bohr.[1]

Nevertheless, the role of aesthetics in the development of quantum physics, and of science in general, has only been addressed in a still rather small literature, in which the monograph by James McAllister on beauty and revolution in science is a key reference.[2] A central theme in McAllister's book and the subsequent literature is whether beauty did, or ought to, influence scientists when evaluating theories. This is famously, and controversially, exemplified by the lesson Paul Dirac inferred from Erwin Schrödinger's development of his wave equation. Dirac explained that Schrödinger's work was guided by a search for mathematical beauty, but also that his equation, by not including electron spin, did not fit the experimental data. This initially kept Schrödinger from publishing his result, and Dirac commented: "I think there is a moral to this story, namely that it is more important to have beauty in one's equations than to have them fit experiment."[3]

But aesthetic considerations in physics are not just about theory evaluation. They may be present at all stages of the scientific process—for instance, in connection with initial motivation (e.g., in the choice of research problem), heuristic guides (e.g., in the choice of research approach), and interpretation in physics. Moreover, aesthetics in physics is not only about beauty and related notions such as symmetry, elegance, and unity. Indeed, aesthetics, in general, is more than beauty, which can be seen from a few historical notes. The word itself stems from the Greek word *aisthesis* (pertaining to the senses), whereas aesthetics as a discipline—concerned with art, beauty and beautiful thinking—was founded in the mid-18th century (by Alexander Baumgarten). In the same century, beauty in nature and art was contrasted with the sublime (by Edmund Burke and later Immanuel Kant), and aesthetics also became related to our play impulse and conceived of as a bridge between the emotional/sensuous and the rational (by Friedrich Schiller). In a scientific context, therefore, a broader conception of aesthetics may include not only beauty but also, for instance, play, the joy of insight (see Section 4), sensuous encounters with natural phenomena, and the notion of the sublime—the latter related to awe, wonder, and that which may be close to the limits of our scientific understanding.[4]

This essay aims to contribute to a better understanding of the different roles of aesthetics in the historical development of quantum physics and of science in general. In the following, I will restrict attention to aesthetic motivation in relation to play, beauty, and the joy of insight. Through a few examples, I will first address the role of these notions in the history of quantum physics, and then consider some contemporary statements from a series of interviews conducted with physicists at the Niels Bohr Institute in Copenhagen. I end with a few considerations regarding the possible dangers of aesthetics in physics.

## 2. Play

Play is an important aspect of science. It appears, for instance, in the sense of puzzle solving, trying out new theoretical ideas, or seeing what happens in unexplored experimental circumstances. The characteristics of play in science include that it is fun, challenging, and closely linked to creativity. Play is also related to aesthetic appreciation insofar as both are what is known in the aesthetic literature as "distinterested". This does not mean that play is uninteresting, but rather that it is enjoyable for its own sake and—at least to some extent—free from the demands of obligations and goals external to the activity itself. Such freedom is a

H. Zinkernagel
Department of Philosophy I, Campus de Cartuja
University of Granada
Granada 18011, Spain
E-mail: zink@ugr.es

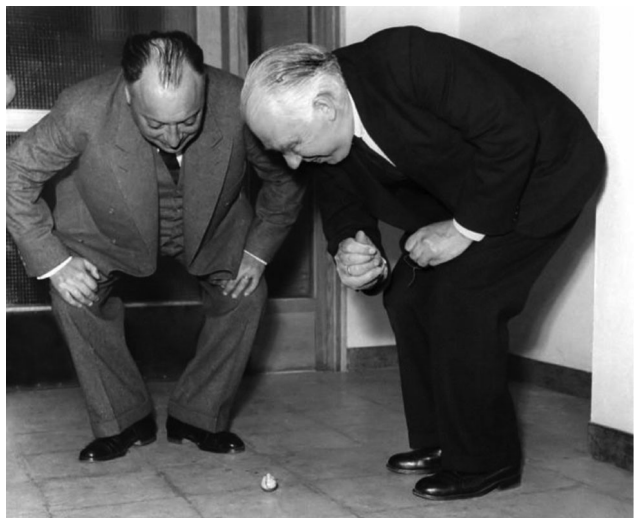

**Figure 1.** Wolfgang Pauli and Niels Bohr playing with a tippe-top in Lund 1951. Reproduced with permission. Copyright The Niels Bohr Archive, Copenhagen.

key element in play, and it is, of course, part of what makes research activity in physics and elsewhere attractive (**Figure 1**). In this way, play can be among the motivations for doing science (see also ref. [5]).[6]

In his autobiography, Victor Weisskopf described the playful spirit at Niels Bohr's institute in Copenhagen in the years from 1924 to 1930, which included the birth of quantum mechanics:[7]

> "…he [Bohr] was always active—talking, creating, living, and working as an equal in a group of young, optimistic, jocu- lar, enthusiastic people who were involved in approaching the deepest riddles of nature with a spirit of attack, freed from conventional bonds in a creative climate that can hardly be described."

Weisskopf's description suggests that playfulness can be an attractive and important part of the *context* in which physics is done (**Figure 2**)—which can, of course, also be seen in the competitive game among physicists of being the first to come up with a new result, or being the one whose theory wins. But there is also an *intrinsic* element of play in physicists' work. This is where play most clearly becomes an aesthetic notion, insofar as it involves disinterestedness—and thus the, at least temporal or partial, suspension of external goals such as the practical usefulness or the personal benefits of having solved a scientific puzzle or explored new territory.

As Johan Huizinga argued in his *Homo Ludens* (Man the Player) from 1949, play is furthermore connected to aesthetics in that beauty is related to order, and play both demands order (a set of rules which must be obeyed for the play situation to be maintained) and creates order (as when a puzzle is solved): "Into an imperfect world and into the confusion of life it [play] brings a temporary, a limited perfection."[8] Indeed, physics also seeks order (e.g., laws of nature) within rules—for example, the rule that experimental results must be reproducible for others. As we shall see, finding or understanding such order is closely connected to the notions of beauty and the joy of insight. For a brief example of how an intrinsic and aesthetic notion of play may have operated among the quantum pioneers, consider Abraham Pais' description of Dirac's way of working: "…first play with pretty mathematics for its own sake, then see whether this leads to new physics."[9]

## 3. Beauty

The idea of beauty has been used as a guideline in the development of new theories since the times of Pythagoras and Plato. Whereas beauty in physics has been associated with ideas such as symmetry, simplicity, and elegance (see, e.g., refs. [10, 11]), a particularly persistent—and related—notion of beauty has been that of unity or connection between different natural phenomena, and between the theories describing these.

Many, if not all, of the quantum pioneers would have agreed with the idea of beauty as unity. For instance, as Heisenberg explained, "Beauty is the proper conformity of the parts to one another and to the whole."[12] The appeal of this notion of beauty is related to the fact that it also expresses a common conception of scientific understanding. Thus, as Wolfgang Pauli supposedly said in a discussion in the early twenties: "'Understanding' probably means nothing more than having whatever ideas and concepts are needed to recognize that a great many different phenomena are part of a coherent whole."[13]

For an early example of beauty as unity, consider this fragment of a letter from the Swedish theoretical physicist C. W. Oseen to Bohr regarding the old quantum theory:

> "…although I already knew the direction of your thinking as well as some of its results, I was still surprised at one point by the beauty of your result. This was the connection between $h$ and the Balmer–Rydberg constant. As far as one can see, on this point you have gone beyond the region of hypotheses and theories and into that of truth itself. Higher no theorist can reach, and I congratulate you with all my heart."
> Oseen to Bohr, 11 November 1913.[14]

However, although a general adherence to beauty as unity prevailed among the quantum pioneers, there were sometimes differences with respect to what kind of unity or connections exactly was to be sought after. As is well known, Einstein was unhappy with quantum mechanics and its abandonment of determinism and causality. In a letter from 1927, Heisenberg challenged Einstein's position in terms of aesthetic ideals:

> "If I have understood correctly your point of view, then you would gladly sacrifice the simplicity [of quantum mechanics] to the principle of [classical] causality. Perhaps we could comfort ourselves [with the idea that] the dear Lord could go beyond [quantum mechanics] and maintain causality. I do not really find it beautiful, however, to demand more than a physical description of the connection between experiments."
> Heisenberg to Einstein, 10 June 1927.[15]

On the other hand, Einstein was not alone in his demand for beauty beyond connections between experimental results. Indeed, the request for a causal description of quantum processes is closely related to one that is visualizable, and it appears that the absence of beauty in this sense may have formed part of Schrödinger's motivation for developing wave mechanics:

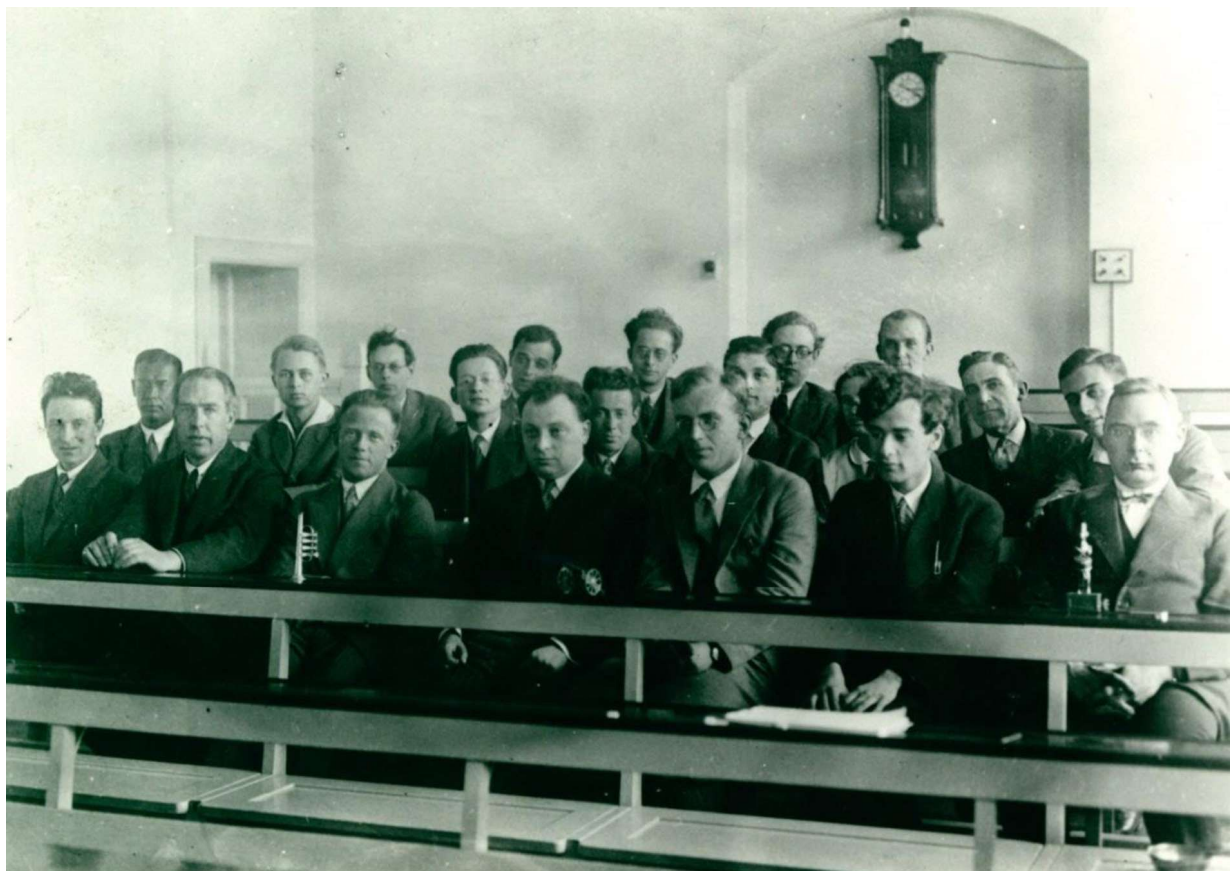

**Figure 2.** Participants at the Copenhagen Conference in 1930. Note the toy trumpet in front of Heisenberg (to be played if the audience liked what was said), the miniature canon in front of Pauli (to be fired if a theory had holes in it), and the toy drummer (in front of Hendrik Kramers) for general applause. Reproduced with permission. Copyright The Niels Bohr Archive, Copenhagen.

> "I naturally knew about his [Heisenberg's] theory, but because of the to me very difficult-appearing methods of transcendental algebra and because of the lack of *Anschaulichkeit* [visualizability], I felt deterred by it, if not to say repelled."
> Schrödinger, May 1926.[16]

## 4. The Joy of Insight

Seeing or discovering beauty in physics is a joyful experience, and if it is accompanied with understanding or a deeper insight into things—e.g., changing the way one sees a theory, an experiment, or some natural phenomena—it will be an example of a joy of insight. Apart from beauty, such joy may involve a sense of wonder or an experience of something sublime, and the joy of insight can more generally be characterized as an aesthetic experience in science. Such experiences appear to be much appreciated, and therefore motivating in physics (and elsewhere). Given their emotional and subjective character, they are harder to document, but they do show up, e.g., in letters, scientific popularization, and autobiographical material.

Consider Heisenberg's telling description of his discovery that the principle of energy conservation comes out right in the new matrix mechanics. This description, of events at the island of Helgoland in June 1925, can be seen as a paradigmatic example of an aesthetic experience in science, which brings out the moving and emotional aspect of a joy of insight:

> "When the first terms seemed to accord with the energy principle, I became rather excited, and I began to make countless arithmetical errors. As a result, it was almost three o'clock in the morning before the final result of my computations lay before me …. At first, I was deeply alarmed. I had the feeling that, through the surface of atomic phenomena, I was looking at a strangely beautiful interior, and felt almost giddy at the thought that I now had to probe this wealth of mathematical structures nature had so generously spread out before me."[17]

From Heisenberg's description (or reconstruction) of this experience, it is clear that more than beauty is involved. In conformity with the aesthetic theory originally laid out by the American philosopher John Dewey, an aesthetic experience may be seen as a "…satisfyingly heightened, absorbing, coherently meaningful and affective experience."[18] These characteristics account for the fact that aesthetic experiences in science are also moving or even transforming, in the sense of changing our perception or view of (aspects of) the world.

As regards Schrödinger, I am not aware of any clear statement on his part regarding a particular joy of insight when he found his wave equation. This may be because Schrödinger's first formulation was the result of a failed attempt to construct an empirically adequate relativistic equation.[19] Other physicists, however, did express their joy over Schrödinger's result in aesthetic terms. Thus, Jürgen Renn (ref. [19b, p.10]) writes:

> "When Max Planck held Schrödinger's second publication in his hands, he related to Schrödinger on a post card: 'I am reading your communication in the way a curious child eagerly listens to the solution of a riddle with which he has struggled for a long time, and I rejoice over the beauties that my eye discovers, which I must study in much greater detail, however, in order to grasp them entirely'."

The joys of insight in physics are (fortunately!) not restricted to those working on the forefront of physics research. They can also be experienced, for instance, by students of physics in their formative years. Here is Weisskopf on this point (ref. [7, p. 35]):

> "As young scientists who had to learn all this, we were overwhelmed by the difficulty of both approaches, although Schrödinger's was easier to grasp by visualizing electrons in the atoms as wavelike vibrations. We were impressed by the fact that these vibrations correspond exactly to the known quantum states of the hydrogen atom. When I began to understand this, I experienced for one of the first times that joy of insight that was to play such a vital role in my life."

Weisskopf's description of this kind of joy in his work conforms to the idea that it is an aesthetic experience. In his autobiography, he explains: "The joy of insight is a sense of involvement and awe, the elated state of mind that you achieve when you have grasped some essential point…" (ref. [7, p. 1]); that it can be (as it was for him in the case of his work on quantum electrodynamics) "…like the opening of a new world" (ref. [7, p. 37]); and that this joy of insight can best be described as "a feeling of aesthetic pleasure" (ref. [7, p. 37]). I think it is safe to assume that such joys of insight—which may involve seeing new connections or the opening of new worlds—can be a source of continuous inspiration in the working life of physicists. Consider, for instance, this remark of Bohr from a 1945 speech:[20]

> "To take part in lifting only a corner of the veil under which truth is hiding and perhaps thereby getting on the track of deeper connections than those immediately apparent, is all the happiness that a researcher can be given."

As the examples of Planck and Weisskopf show, the joys of insight need not be original ones. In this sense, their attractiveness and motivating potential can be shared among physicists, as indicated, for instance, in this description of Bohr by his collaborator Léon Rosenfeld:[21]

> "The voyage of discovery was a meeting between nature and the human intellect, which he experienced with dramatic intensity. Bohr felt the excitement of the investigation and the joy of its completion so intense that he wanted everyone around him to share them with him." [My translation from the Danish]

## 5. Aesthetics in (Quantum) Physics Today

During two research stays in 2016 and 2018 at the Niels Bohr Archive, I conducted a small number (14 in total) of qualitative interviews with senior theorists and experimentalists at the Niels Bohr Institute—and one at the Technical University (with Tomas Bohr, grandson of Niels Bohr)—in Copenhagen.[22] The aim was to get an idea of whether aesthetics still plays a motivating role among physicists who are working, or have worked, with quantum physics at this renowned institute in physics history. Interviewees were asked questions such as "Do you think the idea of play has played any role in your work?", "Has the idea of beauty played any role in your work?", and "Can you relate to Weisskopf's description in his autobiography regarding the 'joy of insight'?"

Since the number of interviews was small and the questions may well have been leading, I can make no claims regarding the generality of the answers. Nevertheless, it was interesting to find that almost all of the interviewees answered "yes" to the three questions listed above.[23] This is consistent with the idea that play, beauty, and the joy of insight are still relevant for aesthetic motivation in today's physics. There is no space here to go into the details of these interviews (all but one are recorded and form part of the collection at the Niels Bohr Archive), but I will briefly summarize some of the findings and provide a couple of quotes. Regarding play, several respondents emphasized its importance due to the relation between play and freedom. Moreover, to my question of whether play was important to her work, Lene Oddershede gave an answer, which nicely illustrates the relations between motivation and play, understanding, and the practical use of physics (in her case, in medical physics):

> "Yes. It must be fun. Otherwise I cannot put in the energy needed for making things work. It is fun for me to be at places where nobody has been before. To find new things, new connections, new characteristics and to understand things. And of course it is also very satisfactory if the things we have discovered can be used in practice and actually make a difference for people."
>
> Interview with Lene Oddershede, 6 December 2018 [my translation from the Danish].

To my question regarding the role of beauty, again many interviewees answered that it was important in their work, and something to strive for either theoretically or in experimental design and results. Tomas Bohr noted that "beauty is involved in the selection of problems [to be studied]". To explain this further, T. Bohr quoted from a written explanation of why certain experiments that he and colleagues had carried out in fluid dynamics on an everyday scale—such as water streaming from a tap or whirling out of a bathtub—were found interesting and even selected for display at an art exhibition:

> "First, these phenomena are sufficiently similar to more violent large-scale natural phenomena, such as tidal waves, tornadoes or galaxies, so that one can learn something significant from them. Second, they pertain to active research areas where new phenomena are constantly being discovered and where there is no agreement on the basic underlying mechanisms. Third, we simply think they are fascinating, which is probably a combination of "beautiful" and "amazing"—without being able to completely define it precisely."
            Interview with Tomas Bohr, 13 December 2018 [my translation from the Danish].

Regarding the question about joy of insight, many of the interviewees recognized the pleasure accompanying such major transcending moments or experiences in the terms described by Weisskopf. Curiously, several respondents replied similarly, when asked for a number, that they had 4–5 of such joys of insight in their careers, even if—as some noted—such experiences of course come in degrees. Among the examples cited were when an experiment carried out in team work eventually works and a specific atomic phenomenon is seen for the first time in history (Jan W. Thomsen), or when an exact mathematical solution to a problem is finally found, leading to an increase in understanding of quantum theoretical structures (Charlotte F. Kristjansen).

## 6. Are there Dangers Associated with Aesthetics?

Even if physicists use aesthetic terms when describing their work, this cannot—of course—establish the extent to which aesthetic considerations actually influenced this work. As noted by Christian Joas (in private communication), physicists might in retrospect "aesthetisize" their work, e.g., by considering it more like an art form or remembering it more beautiful than it actually was. This is most likely true in some cases, but even so, it does not challenge the idea that physics has aesthetic aspects and that physicists respond to these.

There have been more critical voices regarding the role of aesthetics in modern physics. Indeed, Sabine Hossenfelder has argued that a hunch for beauty may be detrimental to the goals of physics—as it can lead physics astray—when physicists (especially in high-energy physics) get too hung up on ideas like "…the now formalized aesthetic ideals of the past: symmetry, unification, and naturalness."[24] I think this is a sound warning. However, as mentioned above, there is more to aesthetics in physics than beauty, so its motivational role is not exhausted by such "ideals of the past." Moreover, these ideals may be understood in different ways. For instance, the striving for unity needs not imply that of theoretical unification but may be limited to establishing connections between different areas. Such is the case, for instance, in Hossenfelder's own research field—quantum gravity—where some work on the unificationist line, aiming for a theory of everything from which all known physics could, in principle, be derived (as in string theory), whereas others work on less ambitious schemes to combine gravitational and quantum physics, see, e.g., ref. [25].

Of course, there is more to motivation in physics than aesthetics—such as advancing knowledge, finding useful applications, or (ambition and) striving for recognition. But I think it is hard to deny that aesthetic motivation has played and continues to play an important role for physicists. Furthermore, aesthetic joy is something which can be shared, e.g., among students, historians, and philosophers of physics. Also for this reason, the continued study of aesthetics in physics seems relevant for many of us.

## Acknowledgements

The author would like to thank Christian Joas, the editors Arianna Borrelli and Tilman Sauer, and Svend E. Rugh for helpful comments to the manuscript. The author also thanks Finn Aaserud for help with the interviews, all the interviewed physicists for sharing their thoughts on these topics, and the Spanish Ministry of Science and Innovation (Project No. FFI2016-77266-P) for financial support. Funding for open access charge (Read and Publish agreement): Universidad de Granada / CBUA.

Note: Minor corrections were made in the text and in refs. [4] and [24], and some typos were fixed, on September 6, 2022, after initial publication online.

## Conflict of Interest

The author declares no conflict of interest.

## Keywords

aesthetics, motivation, quantum physics

of what happened that night in 1925. In any case, and judging also from many other testimonies, there is little reason to doubt that a perceived significant breakthrough in one's scientific understanding will often be accompanied by the type of feelings Heisenberg mentions.

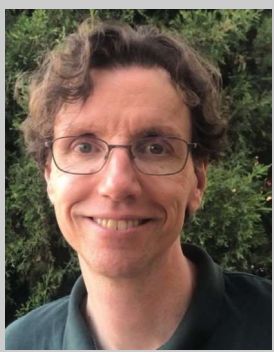

**Henrik Zinkernagel** is Associate Professor of philosophy at the University of Granada. He has a Masters degree in physics from the Niels Bohr Institute and holds a PhD in philosophy of science from the Center for Philosophy of Nature and Science Studies at the Niels Bohr Institute. His work has been centered mainly on the history and philosophy of cosmology, as well as on the history and philosophy of quantum mechanics (especially concerning the views of Niels Bohr). In recent years, his research has focused increasingly on the aesthetic aspects of science and the inclusion of aesthetics in science education.